\documentclass[prd,aps,superscriptaddress,showpacs,amssymb]{revtex4}
\usepackage{epsfig}

\begin{document}

\title{One loop calculation of the renormalized anisotropy for
improved anisotropic gluon actions on a lattice}

\author{I.T. \surname{Drummond}}
\author{A. \surname{Hart}}
\altaffiliation[{Current address: }]{School of Physics, University of
Edinburgh, King's Buildings, Edinburgh EH9 3JZ, U.K.}
\author{R.R. \surname{Horgan}}
\author{L.C. \surname{Storoni}}
\affiliation{DAMTP, CMS, University of Cambridge, Wilberforce Road,
Cambridge CB3 0WA, U.K.}

\begin{abstract}
Using the infrared dispersion relation of the on shell gluon, we
calculate the renormalisation of the anisotropy, $\chi$, to one loop
in perturbation theory for lattice Yang--Mills theories, including the
Wilson action and actions with Symanzik and/or tadpole
improvement. Using twisted boundary conditions as a gauge invariant
infrared regulator, we show for an SU(3) gauge group in $D=3+1$
dimensions that the one loop anisotropy is accurate to $O(3\%)$ for a
range of $g^2$ and $\chi$ covering current simulations. In doing so we
also present Feynman rules for SU(N) gauge groups with generic
anisotropy structure (including `$3+1$' and `$2+2$' cases) for both
twisted and untwisted boundary conditions.
\end{abstract}

\preprint{Cambridge DAMTP-2002-92}


\maketitle



\def\metric{\mathfrak{g}}

\def\bacol{\setlength{\arraycolsep}{0pt}}
\def\bec{\begin{center}}
\def\enc{\end{center}}
\def\[{$$}
\def\]{$$}
\def\ben{\begin{equation}}
\def\ba{\begin{array}}
\def\bea{\begin{eqnarray}}
\def\een{\end{equation}}
\def\eea{\end{eqnarray}}
\def\ea{\end{array}}
\def\btab{\begin{table}}
\def\btabu{\begin{tabular}}
\def\etab{\end{table}}
\def\etabu{\end{tabular}}
\def\bit{\begin{itemize}}
\def\eit{\end{itemize}}
\def\bef{\begin{figure}[htb]}
\def\befh{\begin{figure}[!h!]}
\def\enf{\end{figure}}
\def\sep{[2truemm]}
\def\chr{\langle r^2 \rangle}
\def\tr{\mbox{Tr}}
\def\la{\langle}
\def\ra{\rangle}
\def\pd{\partial}
\def\bbeta{{\beta_0}}
\def\bchi{{\chi_0}}
\def\bg{{g_0}}
\def\a{\alpha}
\def\r{\rho}
\def\hY{\hat{Y}}
\def\hk{\hat{k}}
\def\cR{{\cal R}}
\def\cZ{{\cal Z}}
\def\cS{{\cal S}}
\def\cP{{\cal P}}
\def\cT{{\cal T}}
\def\cU{{\cal U}}
\def\cW{{\cal W}}
\def\cL{{\cal L}}
\def\barA{{\bar{A}}}
\def\barU{{\bar{U}}}
\def\hatphi{{\hat{\phi}}}
\def\tphi{{\tilde{\phi}}}
\def\tp{{\tilde{p}}}
\def\tq{{\tilde{q}}}
\def\tk{{\tilde{k}}}
\def\barD{{\bar{\D}}}
\def\balpha{\mbox{\boldmath $\alpha$}}
\def\bmu{\mbox{\boldmath $\mu$}}
\def\bDel{\mbox{\boldmath $\Delta$}}
\def\bphi{\mbox{\boldmath $\phi$}}
\def\bsig{\mbox{\boldmath $\sigma$}}
\def\bareta{{\bar{\eta}}}
\def\gb{\beta}
\def\gU{\Upsilon}
\def\D{\Delta}
\def\d{\delta}
\def\e{\epsilon}
\def\g{\gamma}
\def\L{\Lambda}
\def\om{\omega}
\def\bLambda{{\mbox{\boldmath $\Lambda$}}}
\def\l{\lambda}
\def\C{\Xi}
\def\O{\Omega}
\def\Pf{\psi_{nlm\s}}
\def\bpsi{\bar{\psi}}
\def\tpsi{\tilde{\psi}}
\def\btpsi{\bar{\tilde{\psi}}}
\def\S{\Sigma}
\def\s{\sigma}
\def\Gm{\tilde{G}}
\def\t{\tau}
\def\dsl{\not\!\d}
\def\G{{\cal G}}
\def\Ga{\Gamma}
\def\GaR{{\Gamma_{\!\! R}}}
\def\U{{\cal U}}
\def\Z{{\cal Z}}
\def\half{{\textstyle{1 \over 2}}}
\def\shalf{{\scriptstyle{1 \over 2}}}
\def\ihalf{{\textstyle{i \over 2}}}
\def\ohalf{{1 \over 2}}
\def\third{{\textstyle{1 \over 3}}}
\def\quarter{{\textstyle{1 \over 4}}}
\def\squarter{{\scriptstyle{1 \over 4}}}
\def\eigth{{\textstyle{1 \over 8}}}
\def\tthirds{{\textstyle{2 \over 3}}}
\def\thalves{{\textstyle{3 \over 2}}}
\def\b1{{\bf 1}}
\def\aI{{\overline I}}
\def\aK{{\overline K}}
\def\aL{{\overline \L}}
\def\tA{\tilde{A}}
\def\bAL{{\overline \bL}}
\def\bALambda{{\overline \bLambda}}
\def\au{{\overline u}}
\def\ad{{\overline d}}
\def\as{{\overline s}}
\def\ar{{\overline r}}
\def\ag{{\overline g}}
\def\ab{{\overline b}}
\def\aq{{\overline q}}
\def\ua{\uparrow}
\def\da{\downarrow}
\def\bc{{\bf c}}
\def\bk{{\bf k}}
\def\bx{{\bf x}}
\def\by{{\bf y}}
\def\bz{{\bf z}}
\def\bp{{\bf p}}
\def\bq{{\bf q}}
\def\be{{\bf e}}
\def\b0{{\bf 0}}
\def\bark{{\bar{k}}}
\def\bme{\mbox{\boldmath $e$}}
\def\bv{\mbox{\boldmath $v$}}
\def\bu{\mbox{\boldmath $u$}}
\def\bn{{\mbox{\boldmath $n$}}}
\def\sbl{{\mbox{\boldmath $\scriptstyle l$}}}
\def\sbm{{\mbox{\boldmath $\scriptstyle m$}}}
\def\sbn{{\mbox{\boldmath $\scriptstyle n$}}}
\def\sbr{{\mbox{\boldmath $\scriptstyle r$}}}
\def\bm{{\mbox{\boldmath $m$}}}
\def\bl{{\mbox{\boldmath $l$}}}
\def\br{\mbox{\boldmath $r$}}
\def\brp{\mbox{\boldmath $r$}^\prime}
\def\bj{\mbox{\boldmath $j$}}
\def\bJ{\mbox{\boldmath $J$}}
\def\bS{\mbox{\boldmath $S$}}
\def\bL{\mbox{\boldmath $L$}}
\def\bT{\mbox{\boldmath $T$}}
\def\bQ{\mbox{\boldmath $Q$}}
\def\bM{\mbox{\boldmath $M$}}
\def\bA{\mbox{\boldmath $A$}}
\def\bB{\mbox{\boldmath $B$}}
\def\bbeta{{\beta_0}}
\def\bchi{{\chi_0}}
\def\bg{{g_0}}
\def\barU{{\bar{U}}}
\def\barM{\bar{M}}
\def\barQ{\bar{Q}}
\def\bart{\bar{t}}
\def\barp{\bar{\mbox{\boldmath $p$}}}
\def\cos{\hbox{cos}\:}
\def\sin{\hbox{sin}}
\def\cot{\hbox{cot}}
\def\cosh{\hbox{cosh}}
\def\sign{\hbox{sign}}
\def\c{\cite}
\def\nn{\nonumber}
\def\bb{\left(}
\def\eb{\right)}
\def\bs{{\bf s}}
\def\br{{\bf r}}
\def\1{{1}}
\def\mod{\mbox{mod}}
\def\pr{\prime}
\def\MS{{\overline{MS}}}
\def\bD{{\bf D}}
\def\U{{\bf U}}
\def\W{{\bf W}}
\def\T{{\bf T}}
\def\bs{{\bf s}}
\def\ss{\scriptscriptstyle}
\def\phip{{\phi^\prime}}
\def\m{m_\infty}
\def\u{\tilde{u}}
\def\ub{\bar{u}}
\def\up{\raisebox{3mm}}
\def\dn{\raisebox{-3mm}}
\def\circf{\begin{picture}(5,5)\put(3,3){\circle*{5.0}}\end{picture}}
\def\circo{\begin{picture}(5,5)\put(3,3){\circle{5.0}}\end{picture}}
\def\sq{\rule{2mm}{2mm}}
\def\rw{\rule[-5mm]{0mm}{12mm}}
\def\rw0{\rule[0mm]{0mm}{15mm}}
\def\rb{\raisebox{3mm}[0pt]}
\def\rb0{\raisebox{0mm}[0mm][20truemm]{\null}}
\def\m{\mbox}
\section{\label{introduction}\bf Introduction}

Lattice Monte Carlo simulations operate by dividing a finite volume of
space and time into a grid, such that in a given direction, $\mu$, we
have $L_\mu$ points distance $a_\mu$ apart. The desire to obtain
results free from uncontrolled finite volume contamination dictates
that the product $a_\mu L_\mu$ be chosen to be suitably large in
spatial directions (3~fm is often quoted for QCD). Controlling
discretization effects similarly requires that $a_\mu$ be suitably
small, but this must be balanced with the computational overhead that
increases with $L_\mu$. Reducing the dependence of simulation results
on the lattice spacing is clearly advantageous, and so--called
(Symanzik) improved actions may achieve this, permitting the use of
coarser lattices without increasing discretization effects
\cite{lepage96}. 

In many cases, lattice results (such as hadron masses or decay
constants) are obtained from the decay of correlation functions,
$C(\tau)$, over a range of temporal separations, $\tau$. It is a
feature of such correlation functions that the signal to noise ratio
decreases with increasing $\tau$, and beyond some $\tau_c$
measurements are dominated by statistical fluctuations. The precise
value of $\tau_c$ depends upon many factors, including the operators
correlated and the number of Monte Carlo measurements made, but
appears to be relatively insensitive to the temporal lattice spacing,
$a_t$.  As measurements can only be made for $\tau$ an integer
multiple of $a_t$, if the temporal lattice spacing is large compared
to $\tau_c$ it will be hard to obtain an accurate picture of $C(\tau <
\tau_c)$.  Improving the action does not help in this respect, and in
addition, the inclusion of improvement in the temporal direction leads
to the introduction of spurious (sometimes called `ghost') poles in
the gluonic propagator
\cite{lepage96}. 
By not temporally improving the action we avoid this, but at the cost
of increased discretization errors for given $a_t$. Controlling these,
and the desire for increased temporal resolution of correlation
functions, argues for the use of a small $a_t$.

We are thus motivated to choose a temporal lattice spacing that is
smaller than the spatial, $a_s$, and such `anisotropic' lattices can
be created by tuning action couplings in the temporal direction
differently to those in the spatial
\cite{karsch82}. 
Anisotropic actions have already been successfully applied in many
situations, including the glueball spectrum
\cite{MP}, 
the spectrum of excitations of the inter-quark potential
\cite{JKM}, 
heavy hybrids 
\cite{maea1,drea0},
and the fine structure of the quarkonium spectrum 
\cite{maea0}.  
More recently, anisotropic lattices have been shown to be very
successful in nonrelativistic QCD (NRQCD) studies of two-- and
three--point correlators and finite momentum hadrons and semileptonic
$B$ decays
\cite{shigemitsu01,shigemitsu02}.

A more widespread use has been hampered by the fact that the bare
anisotropy (or aspect ratio) in the simulated action, $\chi =
a_s/a_t$, is not, due to quantum mechanical effects, the same as the
measured value, $\chi_R$. Typically, we wish to estimate the continuum
limit ratio of a mass, $M$, to a given scale, $K$, using the lattice
measurements, $\hat{M}$ (in units of $a_t$) and $\hat{K}$ (often
derived from the static quark potential, and in units of $a_s$)
respectively. Up to finite lattice spacing corrections,
\ben 
\frac{M}{K} = \frac{\hat{M}}{\chi_R \hat{K}} \; .
\label{eqn_setscale}
\een
We thus require $\chi_R$, and with a sufficiently small error that
this does not represent the dominant uncertainty in the final
estimate. $\chi_R$ may be measured in Monte Carlo simulations, {\it
e.g.}
\cite{alea},
but it is an expensive calculation which must be repeated for every
choice of the bare couplings.

More generally, (lattice) perturbation theory may be used to calculate
the renormalisation of quantities, and it is well known that with
`tadpole improvement' such calculations converge quickly to the
measured data at simulated values of the gauge coupling, $g^2$,
\cite{lepage96}. 
In this paper we obtain $Z(g^2,\chi) \equiv \chi_R/\chi$ to one loop
for a wide range of anisotropies for SU(3) gauge theories in four
dimensions. We focus on the Wilson action and a commonly used Symanzik
improved action. Our results apply both to actions with and without
tadpole improvement; in the Wilson case we cover both plaquette and
(Landau) mean link improvement, whilst in the Symanzik improved case
we discuss here only the mean link improved case.

These is no reason, {\it a priori}, why the lattice spacing should be
identical in all spatial directions, and indeed there are situations
where we might choose this not to be the case. A typical example is
where increased momentum resolution is desired for correlation
functions of operators at finite momentum. Rather than increase the
computational overhead by a global rescaling, $a$ can be made smaller
in one spatial direction
\cite{burgio02}. 
Whilst we do not consider calculations for this case explicitly in this
paper, when we describe the Feynman rules in Section~\ref{sec_pert},
we allow for arbitrary anisotropic lattice structure as well as
general actions. We use twisted boundary conditions as a gauge
invariant infrared regulator. In Section~\ref{sec_anis} we describe
our calculation of the anisotropy from the dispersion relation of the
on shell gluon propagator. We compare these one loop results to
measurements of $\chi_R$ from simulations, and show that the one loop
result is accurate to within 3--4\% over the range of couplings
covered by Monte Carlo simulations. Finally, in
Section~\ref{sec_summary} we provide a summary of our findings and
some conclusions.

\section{\label{sec_pert} The Perturbation Theory}

It is useful to consider the derivation of an anisotropic lattice
action from the isotropic continuum theory in two stages. We first
obtain an action for an anisotropic continuum, which is then
discretized.

\subsection{\label{anisotropy} continuum anisotropy}
\label{subsec_anis}

The starting point, and the fixed point of the lattice action, is a
$D$--dimensional continuum field theory in an Euclidean space-time
that is invariant under Lorentz transformations and hence
isotropic. We may choose to change our measurement units in the
continuum theory, and by different factors in different directions,
which leads to the introduction of an anisotropy factor (or factors),
$\chi$, into the action, being the ratio of the length units in
different directions. Nonetheless, Ward identities (derived by
considering anisotropies differing infinitesimally from unity) can be
enforced to ensure that the underlying theory maintains the correct
Lorentz invariance under renormalisation.

We distinguish quantities in the isotropic theory from those in the
anisotropic by the use of `hats' in the former case. Although the
original metric is
\ben
\hat{\metric}_{\alpha \beta} = {\rm diag}(+1,+1,...,+1), 
\een
we find it convenient to introduce the notation of covariant and
contravariant indices. The contraction of a momentum and position
\footnote{The Einstein summation convention is understood to apply unless
otherwise stated.},
$\hat{\bp} \cdot \hat{\bx} \equiv \hat{p}_\alpha \hat{x}^\alpha$, must
be invariant under rescaling, {\it i.e.}
\ben
\hat{p}_\alpha \hat{x}^\alpha = p_\mu x^\mu.
\label{eqn_invariance}
\een
We can relate the rescaled fields to the original
by the use of a set of vierbeins
\ben
x^\mu = {e^\mu}_\alpha \hat{x}^\alpha \; , ~~~~
p_\mu = {e_\mu}^\alpha \hat{p}_\alpha \; .
\een
The metric in these variables is
\ben
\metric_{\mu \nu} = {e_\mu}^\alpha {e_\nu}_\alpha \; , ~~~~
\metric \equiv \det \metric_{\mu \nu} \; .
\label{eqn_metric}
\een
In the most general case there will be $D-1$ anisotropy factors, but
for the rescaling in the temporal direction only, as we consider in
this paper,
\ben
{e^\mu}_\alpha = \left( \chi,1,...,1 \right) \; , ~~~~
{e_\mu}^\alpha = \left( \frac{1}{\chi},1,...,1 \right) \; ,
\een
and 
\ben
\metric_{\mu \nu} = {\rm diag}
\left( \frac{1}{\chi^2},1,...,1 \right) \; , ~~~~
\metric = \frac{1}{\chi^2} \; .
\een
Using these conventions, the natural position vector is covariant
under rescaling, $\hat{\bx} = (\hat{x}^\alpha) = (t,{\underline x})$,
so the momentum must be contravariant $\hat{\bp} = (\hat{p}_\alpha) =
(E,{\underline p})$ such that Eqn.~(\ref{eqn_invariance}) is
satisfied.  The volume element is given by
\ben
d \hat{\bx} = d \bx \prod_{\mu=1}^{D} {e_\mu}^\alpha
= d \bx \sqrt{\metric} = \frac{d \bx}{\chi}.
\een
The dimensionfull (colour) vector potential and derivatives behave as
$\hat{\bp}$, and so the Yang--Mills action becomes
\ben
S_{\rm YM} =  \frac{1}{4} \int d \hat{\bx} 
\hat{F}_{\alpha \beta} \hat{F}^{\alpha \beta}
= \frac{1}{4} \int d \bx \sqrt{\metric} \:
F_{\alpha \beta} F^{\alpha \beta}
= \frac{1}{4} \int d \bx \sqrt{\metric} \:
\metric^{\mu \sigma} \metric^{\nu \tau}
F_{\mu \nu} F_{\sigma \tau} \; .
\een
For the specific example above,
\ben
S_{\rm YM} = \frac{1}{4} \int \frac{d \bx}{\chi} 
\left( {\chi}^2 F_{0i} F_{0i} + F_{ij} F_{ij} \right) \; .
\label{eqn_anis_ac}
\een

\subsection{\label{prop} the propagator}

To construct the Feynman diagrams for any action the gluon propagator
must be computed. This is done for a given momentum by inverting the
two-point gluon vertex, which for the isotropic (continuum) case is
given by $\hat{V}^{(2) \alpha\beta}(\hat{\bk}) = \hat{k}_\epsilon
\hat{k}^\epsilon \delta^{\alpha \beta} - \hat{k}^\alpha
\hat{k}^\beta$. Before this can be done the gauge must be fixed and we
add to the action a gauge fixing term and source, which in momentum
space appear as
\ben
S_{\rm g.f.} = 
\int \frac{d \hat{\bk}}{(2\pi)^4} \: \frac{1}{2 \a} 
\hat{A}_\mu(\hat{\bk}) \hat{k}^\mu \hat{k}^\nu \hat{A}_\nu(-\hat{\bk}) -
\hat{J}^\mu(\hat{\bk}) \hat{A}_\mu(-\hat{\bk}) \, .
\een
The parameter $\a$ is the usual gauge-fixing parameter and, for
example, $\alpha = 1$ corresponds to Feynman gauge. In moving to the
anisotropic theory, $d \hat{\bk} = d \bk / \sqrt{\metric}$, which
affects functional derivatives with respect to the anisotropic source,
$J^\mu$. We can rescale $A_\mu$ to absorb this metric factor, which
multiplies terms quadratic in $A_\mu$ by $\sqrt{\metric}$. The two
point function, $V^{(2)\mu\nu}(\bk)$, that we shall shortly derive
from the action, will already contain this factor and the inverse
propagator becomes
\ben
\Ga^{(\a)\mu\nu}(\bk) = V^{(2)\mu\nu}(\bk) + 
\frac{k^\mu k^\nu}{\a}  \sqrt{\metric} \; .
\label{inv_bare_prop}
\een
By illustration, the inverse propagator in the continuum for
$\alpha=1$ has the form
\ben
\Gamma^{(\alpha=1)\mu \nu}(\bk) = \delta^{\mu \nu} \sqrt{\metric} 
k_\rho k^\rho =
\frac{\delta^{\mu \nu}}{\chi} \left( 
\chi^2 {\hat{k}_0}^2 + \sum_i {\hat{k}_i}^2
\right)
\label{eqn_inv_cont_prop}
\een
where the latter expression uses the isotropic momenta, which for a
lattice theory we shall equate to the `physical' ones.

The propagator is
\ben
G^{(\a)}_{\mu\nu} = \left( {\Ga^{(\a)\mu\nu}} \right)^{-1}\;.
\label{bare_prop}
\een
To fix to Landau gauge we must be more careful. Consider the case
where we wish to change the gauge from $\a$ to $\g$ after
inversion. Then
\ben
\Ga^{(\g) \mu \nu} = \Ga^{(\a) \mu \nu} + 
\left( \frac{1}{\g} - \frac{1}{\a} \right) 
k^\mu k^\nu\sqrt{\metric} \; .
\een 
We write
\ben
G^{(\g)}_{\mu\nu} = G^{(\a)}_{\mu\nu} + D_{\mu\nu} \; ,
\een
and then $D_{\mu\nu}$ satisfies
\ben
\bb \Ga^{(\a) \mu \nu} +
\left( \frac{1}{\g} - \frac{1}{\a} \right) 
k^\mu k^\nu\sqrt{\metric} \eb
\bb G^{(\a)}_{\nu \r} + D_{\nu \r} \eb = {\delta^\mu}_\r \; .
\een
The solution for $D_{\mu\nu}$ is
\ben
D_{\mu\nu} = - \frac{(\a-\g)}{\sqrt{\metric}}
{k_\mu k_\nu \over (k_\alpha k^\alpha)^2} \; .
\label{eqn_gauge_corr}
\een

\subsection{discretization}

The anisotropically formulated theory may be discretized in the normal
way, and in these anisotropic units we set the lattice spacing, $a$, to be
the same in each direction.

On a cubical lattice $\L_D$ in dimension $D$ (with $\bx \in \L_D$,
$\mu=1,\ldots,D$) the gauge field is denoted $U$, 
\ben
U = \left\{ U_\mu(\bx):~U_\mu(\bx) \in SU(N) \right\} \; ,
\een
where $U_\mu(\bx)$ is associated with the link $(\bx, \mu)$.

We define the perturbative gauge field $A$ by
\ben
A = \left\{ A_\mu(\bx+\half\be_\mu): ~ A_\mu(\bx+\half\be_\mu) 
\in alg(SU(N)) \right\} \; ,
\een
with $\be_\mu$ the lattice basis vectors, all of unit length and
changing covariantly with rescaling. Expanding in the colour index,
$a$,
\ben
A_\mu = A_\mu^a\,T_a \; ,
~~~[T_a,T_b]=-f_{abc}T_c \; ,
~~~\tr(T_aT_b)=-\half\,\d_{ab} \; ,
\een
where $T_a$ are the (anti-hermitian) generators of $SU(N)$ with
structure constants $f_{abc}$. It is expedient to associate the gauge
potential with the centre of the link, and then
\ben
U_\mu(\bx) = \exp(g A_\mu(\bx+\half\be_\mu)) \; ,
\label{ufield}
\een
where $g$ is the bare coupling constant, and we have absorbed a factor
of $a$ into each component of $A_\mu$.

For a lattice with $L_\mu$ sites in the $\mu$ direction the
momentum vector $\bk$ is
\ben
\bk = \frac{2\pi}{a} \, 
({\bar{k}_1 \over L_1}, \ldots, {\bar{k}_D \over L_D}) \; ,
~~
0 \le \bar{k}_\mu < L_\mu \; , 
~~
\bar{k}_\mu \in \mathbb{Z} \; , 
\een
and the sum over $\bk$ stands for the sum over the components
$\bar{k}_\mu$. In the limit that $L_\mu \rightarrow \infty$ we have
\ben
\frac{1}{a L_\mu} \sum_{\bar{k}_\mu} ~ \rightarrow ~ 
\int \frac{dk_\mu}{2\pi} \; .
\een
The Fourier transform to momentum space is
\bea
\tA_\mu(\bk) &=& \sum_\bx\, e^{-i\bk\cdot(\bx+\ohalf\be_\mu)}
A_\mu(\bx+\half\be_\mu),\nn\\ A_\mu(\bx+\half\be_\mu)
&=&{1 \over
V}\sum_\bk\,e^{i\bk\cdot(\bx+\ohalf\be_\mu)}\tA_\mu(\bk) \; ,
\eea
where $V=\prod_\mu L_\mu$ is the number of lattice points. It is
useful to re-express the position of a gauge potential as
$\bv_\mu = 2(\bx+\half\be_\mu)$, which is a $D$--dimensional vector
with integer (covariant) components.

\subsection{vertex functions}

To permit us to compute perturbation theory for a range of actions, we
have developed an algorithmic method for expanding a general gauge
theory action on a lattice in an appropriate form for perturbative
calculations to be carried out. The approach follows closely the
method and notation of L\"{u}scher and Weisz
\cite{luwe} 
but is extended to accommodate, {\it inter alia}, anisotropy, fermionic
actions, actions for non-relativistic heavy quarks (NRQCD) and more
complicated definitions of the action in the purely gluonic
sector. The algorithm is implemented in the {\sc Python} programming
language. For the work presented in this paper we briefly review the
notation relevant to the present calculation and refer to
\cite{luwe}
for further information.

The lattice action for the pure glue sector can be written as a sum
over contours
\ben
S[U] = \sum_{\bx,\a}~c_\a \: \mathrm{Re} \: \tr \; (P_\a(U;\bx)) \; ,
\een
which is defined in terms of the coupling constants, $c_\a$, and the
$P_\a$, which are closed Wilson loops.

The perturbative action is the expansion of $S[U]$ as a
polynomial in $A$ and the coefficients of the monomials will
determine the vertices of the theory. We denote this action as
$S[A]$ and following 
\cite{luwe}
we write
\ben
S[A] = \sum_r{g^r \over r!} \;
\sum_{\bk_1,\mu_1,a_1} \ldots \sum_{\bk_r,\mu_r,a_r}
\tA^{a_1}_{\mu_1}(\bk_1) \ldots \tA^{a_r}_{\mu_r}(\bk_r) \;
V_r(\bk_1,\mu_1,a_1; \, \ldots; \, \bk_r,\mu_r,a_r) \;.
\een
By a choice of units we set the lattice spacing to $a=1$. The value of
the lattice spacing in physical units is determined by a calculation
of a physical dimensionfull quality and depends on $g$ and hence on
the renormalized coupling constant through the standard $\gb$
function. Other quantities, such as the bare anisotropy, are
determined by the coupling coefficients $c_\a$.

The Euclidean Feynman rule for the $r$-point gluon vertex function is
$(- g^r V_r)$, where the vertex $V_r$ can be expressed as
\cite{luwe}
\ben
V_r(\bk_1,\mu_1,a_1; \, \ldots ; \bk_r,\mu_r,a_r)
= \sum_{\s \in \cS_r}
\s \cdot C_r(a_1,\ldots,a_r) ~ 
\s \cdot Y_r(\bk_1,\mu_1; \, \ldots ; \bk_r,\mu_r) \; ,
\een
where we symmetrize over
$\cS_r$, the permutation group of $r$ objects.

The $C_r$ are the Clebsch-Gordan coefficients which, owing to the
reality of the action, are defined by
\ben
C_r(a_1,\ldots,a_r) = \tr \; ( T_{a_1} \ldots T_{a_r}) + 
(-1)^r \tr \; ( T_{a_r} \ldots T_{a_1}) \; .
\label{cg_notwist}
\een
Under $\cZ_r$, the subgroup of cyclic permutations and inversion, the
$C_r$ have simple properties,
\bea
\s \cdot C_r & = & \chi_r(\s)\;C_r~~~~\s \in \cZ_r,\nn\\
\chi_r(\s) & = & \left\{ 
\begin{array}{l}
1~~~\mbox{for~~} \s \mbox{~~a cyclic permutation}, \\
(-1)^r~~~\mbox{for~~} \s \mbox{~~the inversion,}
\end{array}
\right.
\eea
so it is useful to split the symmetrization operation into two steps
\bea
V_r(\bk_1,\mu_1,a_1; \, ... ; \,\bk_r,\mu_r,a_r)
& = & \sum_{\s \in \cS_r/\cZ_r}
\s \cdot C_r(a_1,... ,\a_r) ~ 
\s \cdot Y_r(\bk_1,\mu_1;\, ... ;\,\bk_r,\mu_r) \; ,
\nn \\
Y_r & = & \sum_\a Y^\a_r\;,~~~~
Y^\a_r = \sum_{\s \in \cZ_r} \; \chi_r(\s) \: \s \cdot \bar{Y}^\a_r \; .
\label{v_notwist}
\eea
The symmetrization over $\cZ_r$ is carried out within the {\sc Python}
vertex generation code, whereas any remaining permutations (for $r
\ge 4$) must be carried out during the loop integration.
$\bar{Y}^\a_r$ is the contribution from the Wilson loop $P_\a$ given
by a sum of $n_r$ terms with the same momentum and Lorentz structure
\ben
\bar{Y}^\a_r(\bk_1,\mu_1;\ldots;\bk_r,\mu_r) = c_\a\sum_{n=1}^{n_r}
\frac{f^\a_n}{2r!}  \; \exp \left( \: \frac{i}{2} \left( 
\bk_1\cdot\bv^n_1 + \ldots +
\bk_r\cdot\bv^n_r \right) \right) \; .
\label{eqn_y}
\een
The factor of $(r!)$ normalises the symmetrization, and the
dependence on the Lorentz indices $\mu_1,\ldots,\mu_r$ has been
suppressed. The prefactor of $\frac{1}{2}$ normalises
Eqn.~(\ref{cg_notwist}). The expansion of $S[U]$ can thus be
represented as a set of `entities' $[f_n;\bv^n_1,\ldots,\bv^n_r]$,
$n=1,\ldots,n_r$, where $f_n$ is an amplitude which is an integer for
simple actions. The {\sc Python} code produces data files where these
are appropriately labelled so that, given the Lorentz indices
$(\mu_1,\ldots,\mu_r)$ and the incoming momenta $\bk_1,\ldots,\bk_r$,
the corresponding value of the $r$-point vertex function $V_r$ can be
computed. The relevant Feynman diagrams can be constructed and the
integrals over loop momenta performed either by direct summation over
modes or using numerical integration routines. The gluon $r$-point
functions are generated with the anisotropy fixed at the chosen (bare)
value, and thus encoded in the amplitudes, $f$. We find this allows
greater simplification of the data files produced by the {\sc Python}
and more efficient loop integral evaluation code. The time taken to
rerun the vertex generation code for different $\chi$ is negligible,
especially when offset against this. A more complete description of
the implementation may be found in
\cite{lat02,meanlink02}.

The gluonic propagator is derived as {\it per} the continuum theory,
using the two point vertex for the particular action,
$V_2(\bk,\mu;-\bk,\nu)$, and pairs of forward and backward nearest
neighbour difference operators,
\bea
\hat{\D}_\alpha & = & (\D_0,\D_i) ~~~\Rightarrow
~~~\D_\mu = (\frac{\D_0}{\chi}, \D_i)\;,
~~~\D^\mu = (\chi \D_0, \D_i)\;,\nn
\\
\hat{\D}^2 & = & \hat{\D}^\pm_\alpha \hat{\D}^{\mp \alpha} =
\hat{\D}^\pm_0 \hat{\D}^\mp_0 + 
\sum_i \hat{\D}^\pm_i \hat{\D}^\mp_i \;.
\label{deriv}
\eea
to replace the position space derivatives in the gauge fixing
term. The difference operators are
\bea
\hat{\D}^+ & = & f(\bx + \hat{\bmu}) - f(\bx)
\nn \\
\hat{\D}^- & = & f(\bx) - f(\bx - \hat{\bmu})
\eea
for some $f(\bx)$.
The net effect is merely to replace momentum components,
$p_\mu$, prior to any raising of the index, by $\tilde{p}_\mu \equiv 2
\; \sin (p_\mu/2)$ in
Eqns.~(\ref{inv_bare_prop}--\ref{eqn_gauge_corr}).

\subsection{Faddeev--Popov ghosts}

The Faddeev--Popov ghost term is of the form $S_{\rm ghost} = \int d
\bx \: \bareta^a M^{ab} \eta^b \; $, where $\bareta$ and $\eta$ are
the usual adjoint anti-commuting ghost fields. The ghost fields are
not observable, and form only internal lines in Feynman diagrams. We
are thus free to choose the normalisation of the fields such that
explicit factors of $\sqrt{\metric}$ do not appear in the momentum
space Feynman rules
\footnote{To match the continuum normalisation, these factors may be
reintroduced, being $1/\sqrt{\metric}$ in the ghost propagator and
$\sqrt{\metric}$ in the vertices, as was actually used in the
calculations for this paper.}
for the ghosts.
The anisotropy then only appears implicitly in the raising of indices.

The Faddeev--Popov matrix $\D^{ab}$ is determined by the gauge fixing
condition corresponding to the choice of gauge in the propagator. The
gauge-fixing is done by introducing the identity in the form
\ben
1 = \int d\a\;\int\prod dA_\mu\;\delta(F[A_\mu]-\l)\exp(-\l^2/2\a)\;.
\een
We use the linear gauge function $(F[A_\mu] = {(\D^-)}^\mu A_\mu$ and, as
is well known, the matrix $\D^{ab}$ is independent of $\a$ in this
case. We denote the gauge transformation field by
\ben
g(\bx) = \exp(\om(\bx))\;,
\een
where $\om(\bx) = \om_a(\bx)T^a$. For infinitesimal $\om$ the gauge
field transforms as
\bea
A^{\om\;a}_\mu&=&A^a_\mu+\O^a_\mu-{g \over 2}\barA^{ab}_\mu\O^{+\;b}_\mu~+~ 
\sum_{n=2}^\infty\;g^n\a_n(\barA^n_\mu)^{ab}\cdot\O^b_\mu\;.\nn\\
\O_\mu&=&-\D^+_\mu\om\;,~~~\O^+_\mu = (2+\D^+_\mu)\om\;,\label{a_om}
\eea
where $\barA_\mu$ is the adjoint representation for the gauge field
and $\a_n$ is the coefficient of $x^n$ in the
expansion of
\ben
b(x) = {x \over e^x-1} = \sum_{n=0}^\infty \a_n x^n\;.\label{b_fn}
\een
The Faddeev--Popov matrix is then
\bea
M^{ab}[A_\mu] & = & 
\left.{\delta F[A^{\om\;a}_\mu] \over \delta \om^b
}\right|_{\om=0} 
\nn \\
& = & {(\D^-)}^\mu \left[-\D^+_\mu\delta^{ab}-{g\over
2}\barA^{ab}_\mu(2+\D^+_\mu)
-\sum_{n=2}^\infty\;g^n\a_n(\barA^n_\mu)^{ab}\D^+_\mu\right]\;.
\label{fd_matrix}
\eea
The inverse ghost propagator is given by the $O(g^0)$ term and is
\ben
\D^{ab}(\bx,\by) = -\delta^{ab}\D^2(\bx,\by)\;,
\een
giving rise to the standard $1/p^2$ ghost propagator.

The one-gluon vertex is given by the $O(g)$ term which gives the
contribution to the action
\ben
g\bareta^af^{abc} {(\D^-)}^\mu \bb A^b_\mu(1+\D^+_\mu/2)\eta^c\eb\;,
\een
and, integrating by parts, we get
\ben
-g({(\D^+)}^\mu\bareta^a)f^{abc}A^b_\mu\eta^c -
\half g({(\D^+)}^\mu\bareta^a) f^{abc}(\D^+_\mu\eta^c) \; .
\label{g3}
\een
The first term generates the standard three-point vertex that one
expects from the continuum but the second term is a lattice artifact
which is suppressed by a power of the lattice spacing $a$ as we should
expect.

At order $g^2$ there is a two-gluon vertex which, using $\a_2=1/12$,
can be read from Eqn.~(\ref{fd_matrix}) to be
\ben
{g^2 \over 12}({(\D^+)}^\mu\bareta^a)\:A^d_\mu A^e_\mu\:f^{dac}f^{ecb}\:
(\D^+_\mu\eta^b)\;.
\label{g4}
\een
Higher order vertices follow a similar pattern but do not contribute
to the one loop calculation we are considering.

\subsection{the Haar measure}

The field measure in the function integral is the Haar measure for
integration over the lattice fields $U_\mu(\bx)$ which take values in
the Lie group. For the perturbative calculation this measure is
re-expressed as the measure for the fields $A_\mu(\bx)$, which take
values in the Lie algebra, times a Jacobian which can be expanded
perturbatively and included as counter terms in the perturbative
action. We relate the infinitesimal vectors $d\xi = d\xi_aT^a$ and $dA
= dA_aT^a$ in the fundamental representation of the Lie algebra by
\ben
e^{g\,d\xi} = e^{-gA}\;e^{g\,(A+dA)}
\een
from which we derive the relation
\ben
dA^a = b(g\barA)^{ab}d\xi^b
\een
where again $\barA$ is in the adjoint representation and the function
$b(x)$ is defined in Eqn.~(\ref{b_fn}) above. The Haar measure is
\ben
\prod_{\mu,\bx}d\xi_\mu(\bx) = 
\prod_{\mu,\bx}{\partial \xi_\mu(\bx) \over \partial A_\mu(\bx)}\;
\prod_{\mu,\bx}dA_\mu(\bx)\;.
\een
The Jacobian then leads to the term in the action
\ben
S_{\rm meas} = -\int\: d \bx \: \sum_\mu\:\tr\left[\log 
b(g\barA_\mu(\bx))\right] = 
\sum_{n=0}^\infty\;g^n\:\gb_n\int\: d \bx
\sum_\mu\:\tr\bb\barA_\mu^n(\bx)\eb\;.
\een
Noting that 
\ben
-x{d\over dx}\log b(x) = b(x)\;,
\een
we find that $\gb_n = \a_n/n$ with $\a_n$ the coefficients in the
expansion of $b(x)$ defined in Eqn.~(\ref{b_fn}).

The $O(g^2)$ vertex from the measure relevant to the one loop
calculation is then
\ben
{1 \over 8}\int\: d \bx \:\sum_{a,\mu}\:[A^a_\mu(\bx)]^2 \; .
\een
\subsection{twisted boundary conditions}

We follow L\"{u}scher and Weisz and use twisted periodic boundary
conditions for the gauge field. There is then no zero mode and hence
no concomitant infrared divergences in the gluon self energy whilst
gauge invariance is maintained. We briefly review these boundary
conditions and refer to 
\cite{luwe} 
for further details.

For an orthogonal twist the twisted boundary condition for gauge fields is
\ben
U_\mu(\bx+L_\nu \be_\nu) = \O_\nu U_\mu(\bx)\O^{-1}_\nu\;,
\label{tw}
\een
where the twist matrices $\O_\nu$ are constant $SU(N)$ matrices which
satisfy
\ben
\O_\mu\O_\nu = z_{\mu\nu}\O_\nu\O_\mu
\een
and $z_{\mu\nu}=\exp(2\pi i n_{\mu\nu}/N)$ is an element of the centre of
SU(N) with $n_{\mu\nu} = 0,1, \ldots, N-1$. The particular boundary
conditions imposed are uniquely specified by the antisymmetric integer
tensor $n_{\mu\nu}$ and a complete discussion may be found in
\cite{gonzalez97}. 
The condition that the twist be orthogonal is that $n_{\mu\nu}\tilde{n}_{\mu\nu} = 0~\mod N$, 
where $\tilde{n}_{\mu\nu} = \epsilon_{\mu\nu\sigma\rho}n_{\sigma\rho}$.
The gauge potential $A_\mu(\bx)$ also satisfies the periodicity condition 
in Eqn.~(\ref{tw}).

Following 
\cite{luwe}
we choose $n_{\mu\nu} = 0$ everywhere, save $n_{12} = -n_{21} =
1$. $\O_1$ and $\O_2$ are then determined up to a unitary $SU(N)$
transformation and $\O_3 = \O_4 = \1$. In the case of orthogonal twist 
$\O_3$, $\O_4$ can be expressed in terms of $\O_1$ and $\O_2$ once the values
of $n_{\mu\nu}$ are given. This will affect the momentum spectrum in
the $3,4$-directions, but the Feynman rules given below are unchanged.

The lattice is here taken to be continuous in the $3,4$-directions and
of extent $L$ sites in the $1,2$-directions. The momentum spectrum,
$\bk$, is then continuous in $k_3,k_4$ and discrete in $k_1,k_2$ with
\ben
k_i = {2\pi \bar{k}_i \over L_i} + {2\pi n_i \over NL_i},~~~~~i=1,2\;,
~~~~~
0 \le \bar{k}_i < L_i,~~~~~0 \le n_i < N \; , ~~~~~
\bar{k}_i, n_i \in \mathbb{Z}\;,
\label{spectrum}
\een
with $n_1=n_2=0$ excluded to eliminate the zero mode and impose a
gauge-invariant infrared cutoff momentum of $2\pi/NL$. Negative
momentum in these directions is $-k_i = (-\bar{k}_i,-n_i)$, adding
appropriate multiples of $L$ and $N$ to remain in the ranges defined
above.

The Fourier expansion of a colour field $\phi(\bx)$ is 
\ben
\phi(\bx) = {1 \over N L_1 L_2}\sum_{k_1, k_2}\int\;
{dk_3 \over 2\pi}{dk_4 \over 2\pi}\;
e^{ik(x+\shalf\be_\mu)}\Ga(\bn)\tphi_\sbn(\bk)\;,
\label{ft}
\een
where $\tphi_\sbn(\bk)$ is a scalar field and the sum over $k_1,k_2$
signifies the sum is over $\bar{k}_1,\bar{k}_2$ and the twist vector
$\bn \equiv (n_1,n_2)$. The $N^2-1$ $SU(N)$ matrices, $\Ga(\bn)$, are
given in terms of $\O_1,\O_2$ by
\ben
\Ga(\bn) = z^{(n_1+n_2)(n_1+n_2-1)/2}\O_1^{-n_2}\O_2^{n_1}\;,
\een
where $z = \exp(2i\pi/N)$ is an element of the centre of $SU(N)$.
We do not need to construct the $\Ga(\bn)$ explicitly but only
evaluate the trace algebra associated with the perturbative
vertices. We introduce the coefficients
\ben
\g_{\sbn a} = \tr(\Ga^\dagger(\bn) T_a)\;,
\een
for which we have the relations
\ben
\ba{rcl}
{2 \over N}\g^\dagger_{a\sbn}\g_{\sbn b}&=&\delta_{ab} \; ,
\\
{2 \over N}\g_{\sbn a}\g^\dagger_{a \sbm}&=&\delta_{\sbn \sbm} \; ,
\\
T_a&=&{1 \over N}\Ga(\bn)\g_{\sbn a} \; ,
\\
\Ga(\bn)&=&2 T_a\g^\dagger_{a \sbn} \; .
\ea
\label{g_relations}
\een
In addition, for an adjoint field we can define the set of scalar
fields labelled by $\bn$
\ben
\phi_\sbn(\bx) = \g_{\sbn a}\phi_a ~~ \Rightarrow ~~ \phi(\bx) = 
{1 \over N}\Ga(\bn)\phi_\sbn(\bx)\;.
\label{eqn_adj_tw}
\een
Using Eqn.~(\ref{ft}) $\phi_\sbn(\bx)$ has Fourier transform
$\tphi_\sbn(\bk)$. Note that the related scalar field
\ben
\bar{\phi}_\sbn(\bx) = e^{-2\pi i(n_1x_1+n_2x_2)/NL}\phi_\sbn(\bx)\;,
\een
is periodic on the lattice and has a momentum spectrum defined by the
$\bar{k}_i$ in Eqn.~({\ref{spectrum}) which allows the numerical
Fourier transform to be easily computed.

Defining the symmetric and antisymmetric products of twist vectors
\bea
(\bn,\bm) & = & n_1 m_1 + n_2 m_2 + (n_1 + m_1)(n_2 + m_2) \; , 
\nn \\
\la \bn,\bm \ra & = & n_1 m_2 - n_2 m_1 \; ,
\eea
the Clebsch-Gordan coefficients given in Eqn.~(\ref{cg_notwist}) are
modified to become
\ben
C^{TW}_r(\bn_1,\ldots,\bn_r) = \frac{1}{N} \left(
\tr(\Ga(\bn_1)\ldots \Ga(\bn_r))~+~(-1)^r\tr(\Ga(\bn_r)\ldots \Ga(\bn_1))
\right) \; .
\label{cg_twist}
\een
The $C^{TW}_r$ can be evaluated using the relations
\ben
\ba{rcl}
\Ga(\bn)&=&\1~~~\bn = \b0~\mod~N \; ,
\\
\tr(\Ga(\bn))&=&0~~~\bn \ne \b0~\mod~N \; ,
\\
\Ga(\bn)^\dagger&=&z^{-\shalf(\sbn,\sbn)}\Ga(-\bn) \; ,
\\
\Ga(\bn^\pr)\Ga(\bn)&=&z^{\shalf(\la \sbn^\pr,\sbn\ra-(\sbn^\pr,\sbn))}
\Ga(\bn^\pr+\bn) \; ,
\ea
\een
where $(\bn~\mod~N)$ is understood to apply to each component,
$n_{1,2}$, and the argument of $\Ga$ is evaluated $\mod~N$. We then
derive the useful result
\ben
C^{TW}_r(-\bn_1,\ldots,-\bn_r) = 
(-1)^r z^{-\shalf\sum_{i=1}^r (\sbn_i,\sbn_i)}
{C^{TW}_r}^*(\bn_1,\ldots,\bn_r)\;. 
\label{useful}
\een
For the inverse propagator we have
\ben
C_2^{TW}(\bn,-\bn) = 2 z^{(\sbn,\sbn)/2}\;.
\een
The $r$-point vertex function is then given in a similar form to that
in Eqn.~(\ref{v_notwist}) by
\ben
V_r(\bk_1,\mu_1;\,\ldots;\,\bk_r,\mu_r) = 
\sum_{\s \in \cS_r/\cZ_r}\;
 \s \cdot C^{TW}_r(\bn_1,\ldots,\bn_r)\;
\s \cdot Y_r(\bk_1,\mu_1;\,\ldots;\,\bk_r,\mu_r)\;.
\label{v_twist}
\een
Note that the momentum arguments $\bk$ implicitly define the
associated twist integers~$\bn$. To simplify the notation, we replace
in most subsequent expressions the twist vector with its `parent'
momentum, understanding that only the twist vector will contribute in
functions such as $\Gamma(\bk)$.

The structure of the vertex functions $Y_r$ is unaffected by the
choice of boundary condition which is manifested only in the momentum
spectrum used.

A simplifying feature is to note that all diagrams contributing to an
$r$-point function carry the same overall phase factor from the centre
of the gauge group.  These phases can be taken out as overall factors
and the remaining parts of the Clebsch-Gordan coefficients and
propagator are real. The overall phase can be restored at the end of
the calculation.

For the 3-point vertex on the left hand side of
Fig.~\ref{prop_graphs}(a) we have (using momentum conservation
$\bp+\bk+\bq = 0$)
\footnote{We use $(123)$, $(1234)$ etc. as a shorthand for the
ordering of the legs in the various permutations $\s \in \cS_r/\cZ_r$
of the vertex.}
\bea
C^{TW}_3(123) 
& = & {1 \over N} \tr(\Ga(-\bq-\bp)[\Ga(\bq),\Ga(\bp)] \nn
\\
& = & {1\over N} \tr(\Ga(\bq+\bp)\Ga(-\bq-\bp))
\bb z^{\shalf ( \la \bq,\bp \ra - ( \bq,\bp) ) } - 
z^{\shalf ( \la \bp,\bq \ra - ( \bp,\bq) ) } \eb \nn 
\\
& = & 2i \, \sin \bb {2\pi \over 3} \la \bq,\bp \ra \eb 
z^{\squarter( (\bq,\bq) + (\bp,\bp) + (\bq+\bp,\bq+\bp))}.
\eea
In the diagram contributing to the gluon self energy of
Fig.~\ref{prop_graphs}(a) the phase factors for both vertices are
identical and, with the phases of internal lines, yields an overall
phase of $z^{\shalf(\bp,\bp)}$ as expected for a term in the
self energy.

For the 4-point vertex loop of Fig.~\ref{prop_graphs}(b) there are
three contributions from the permutations $S_4/\Z_4$ corresponding to
$(1234)$, $(1243)$, $(1324)$ and in each case the $Y_4$ are real. We
then find the Clebsch-Gordan factors to be
\ben
\ba{rcl}
C_4^{TW}(1234) & = & \ba{rcl} C_4^{TW}(1243) & = & 2 z^{(\bk,\bk)/2}
z^{(\bp,\bp)/2} \ea\\
\\
C_4^{TW}(1324) & =  & 2 \cos \bb {2 \pi \over 3} \la \bk,\bp \ra \eb 
z^{(\bk,\bk)/2}z^{(\bp,\bp)/2},
\ea
\een
The $z^{(\bk,\bk)/2}$ cancels the $z^{-(\bk,\bk)/2}$ from the internal
propagator giving an overall factor of $z^{(\bp,\bp)/2}$ which is the
expected phase.

\subsection{ghost and measure Feynman rules}

We choose the Fourier representation for the anti-ghost field
$\bareta$ given in Eqn.~(\ref{ft}) but use the conjugate twist
matrices $\Ga^\dagger(-\bn)$. In this case the ghost propagator is
real and given by
\ben
\D(\bp,-\bp) = 
\frac{1}{\tp_\alpha \tp^\alpha} \; .
\label{eqn_ghost_prop}
\een
Now consider the relevant part of the $\bareta A \eta$ vertex,
\bea
\int d \bx \: f^{abc} \bareta^a A^b \eta^c & = & 
\int d \bx \: 2 \tr ( \bareta[A,\eta]) \nn
\\  
& = & {2 \over N^3} \tr ( \Ga^\dagger(-\bk) [\Ga(\bp),\Ga(\bq)] )
\int d \bx \: \bareta_\bk A_\bp \eta_\bq \; .
\eea
where terms such as $\bareta_\bk$ refer to the decomposition in
Eqn.~(\ref{eqn_adj_tw}), with the momentum subscript restricted to the
twist vector. The Clebsch-Gordan structure is
\ben
F_3(\bk,\bp,\bq) = 
\tr(\Ga^\dagger(-\bk)[\Ga(\bp),\Ga(\bq)]) = 
z^{-\shalf(\bk,\bk)}C_3^{TW}(\bk,\bp,\bq)
\een
From
the full structure of the $\bareta A\eta$ vertex in Eqn.~(\ref{g3})
the momentum space Feynman vertex with momentum assignment shown in
Fig.~\ref{prop_graphs}(c) is
\ben
-i g \tk^\mu \cos (\frac{q_\mu}{2}) F_3(\bk,\bp,\bq) \; ,
\label{eqn_f3}
\een
where there is no implied sum over $\mu$.  Note that
$C_3^{TW}(\bk,\bp,\bq)$ plays the {\it r\^{o}le} of a structure
constant. With a real ghost propagator this
vertex contributes the same phase factor as does the 3-point gluon
vertex and so it can be absorbed into the overall phase factor of the
diagram.

The two-gluon vertex, Eqn.~(\ref{g4}), can similarly be
analysed. Assigning the ghost momentum ($-\bk$ in
Fig.~\ref{prop_graphs}(d)) as $\bq$, and that of the second gluon
($-\bp$ in Fig.~\ref{prop_graphs}(d)) as $\br$, the Clebsch-Gordan
factor is
{\bacol
\bea
F_4(\bk,\bp,\bq,\br) = 
z^{-\shalf(\bk,\bk)} \sum_{\bl} z^{-\shalf(\sbl,\sbl)}
\left( \right. && C_3^{TW}(\bk,\bp,\bl) C_3^{TW}(\bq,\br,-\bl) 
\nn \\
&& \left. + C_3^{TW}(\bk,\br,\bl) C_3^{TW}(\bq,\bp,-\bl) \right) \; .
\eea 
}
In the above we sum over a twist vector, $\bl$. The corresponding
Feynman vertex is, with no sum implied over $\mu$,
\ben
-\frac{g^2}{12} \tk^\mu \tq_\mu 
\delta_{\mu \nu} F_4(\bk,\bp,\bq,\br) \; .
\label{eqn_f4}
\een
The Lorentz index of the second gluon is~$\nu$. It is not so easy to
extract an overall centre phase for twisted boundary conditions, but
for the contribution to the propagator self energy in
Fig.~\ref{prop_graphs}(d) we use Eqn.~(\ref{useful}) to find the
vertex contribution is proportional to 
$z^{\shalf(\bp,\bp)} \: |C_3^{TW}(\bk,\bp,\bl)|^2$
(where we sum over twist vector $\bl$), which carries the phase
appropriate to the self energy.

The measure creates an insertion in the gluon propagator, and at
leading order in~$g^2$, is
\ben
g^2 \frac{z^{\shalf(\bp,\bp)}}{2} \delta_{\mu\nu} \; ,
\een
which carries the correct phase. As expected, these
expressions are independent of the anisotropy.

Although we do not utilise them in this paper, for completeness we
also give the Feynman rules for untwisted boundary conditions in our
notation. The ghost propagator, Eqn.~(\ref{eqn_ghost_prop}), gains an
extra factor of $\delta_{ab}$ for the ghost colour indices, and the
measure insertion becomes
\ben
-g^2 \frac{\delta_{\mu \nu} \delta_{ab}}{4} 
\een
at leading order. In the vertices of
Eqns.~(\ref{eqn_f3},~\ref{eqn_f4}), the Clebsch-Gordan factors are
replaced by
\bea
F_3(\bk,\bp,\bq) & = &
2 \tr (T_\bk [T_\bp,T_\bq]) = 2 C_3(\bk,\bp,\bq) = 
-f_{\bk \bp \bq} \; ,
\nn \\
F_4(\bk,\bp,\bq,\br) & = &
4  \sum_{e} \left(
C_3(\bk,\bp,e) C_3(\bq,\br,e) + C_3(\bk,\br,e) C_3(\bq,\bp,e) 
\right) \; ,
\eea
where, in $F_4$, we sum over a colour index, $e = 1 \ldots N^2-1$. In
both expressions, each momentum factor is understood to be replaced
by the colour index associated with that leg of the vertex.

\section{\label{sec_anis} Anisotropy Renormalisation}

In a lattice simulation, the renormalized anisotropy is typically
determined by comparing the correlation lengths of an operator
measured along different lattice axes. In a perturbative calculation
there is a much smaller choice of quantities sensitive to the
anisotropy.

The use of twisted boundary conditions provides one such, by providing
a gauge invariant gluon mass. The renormalized anisotropy can be
derived from the calculation of the on-shell dispersion relation for
the gluon propagator defined in the theory with twisted boundary
conditions. The details of the theory are fully discussed by
L\"{u}scher and Weisz in
\cite{luwe} 
and we shall follow their notation. We use a lattice of size $L$ in
the $1,2$--directions to which the twisted boundary conditions apply,
and of extent $I \to \infty$ in the $0,3$--directions. We consider the gluon
mode (called the $A$-meson in 
\cite{luwe}) 
which has (Euclidean) momentum
\ben
\bp = (iE_0,m_0,0,p_3)\;,~~m_0 = {2\pi/NL}\;,
\label{mom}
\een
where $E_0$ and $p_3$ are continuous. In this section we understand
all momentum components to be measured in the same units, i.e. to
refer to the isotropic coordinate system. For clarity of presentation,
however, we omit the carets used to distinguish such quantities in
Section~\ref{subsec_anis}. Infrared divergences are regulated by
finite $L$, and $m_0$ is the pole mass of the bare gluon
propagator. In
\cite{luwe} 
the one loop renormalisation of $m_0$ is calculated and is used to
determine the radiative corrections to parameters in the improved
action. We follow a similar procedure and for anisotropic actions
calculate the pole energy of the propagator as a function of
$p_3$. For sufficiently small $p_3$ the infrared dispersion relation
so derived can be fitted to the standard quadratic form using the
renormalized mass and renormalized anisotropy as parameters.

To carry out the calculation we use Feynman gauge as described in
Section~\ref{prop} with $\a=1$, but we verified that the results were
unchanged when other gauges are chosen, as demanded by gauge
invariance. The diagrams that contribute to the one loop gluon self
energy $\S_{\mu\nu}(\bp)$ are shown in Fig.~\ref{prop_graphs} and the
Feynman integrals were constructed using the vertices and rules of the
previous section.

At tree level the on-shell dispersion relation is given by
$\Gamma^{(\a=1)}_{\mu\nu}(\bp) = 0$ for $\bp$ of the form above, and
where $\Ga^{(\a)}_{\mu\nu}(\bp)$ is defined in
Eqn.~(\ref{inv_bare_prop}).  In the continuum,
Eqn.~(\ref{eqn_inv_cont_prop}), this gives
\ben
\sqrt{\metric} p_\mu p^\mu = \frac{1}{\chi} \left(
-\chi^2 {E_0}^2 + {p_3}^2 + {m_0}^2 \right) = 0
\een
and the bare mass is defined by $E_0 = m_0 / \chi$ at $p_3 = 0$. On
the lattice for very small $p_3$ this becomes
\ben
\frac{1}{\chi} \left( -\chi^2 F_t(E_0) + {p_3}^2 + F_s(m_0) \right) = 0
\label{eqn_tree_zero}
\een
where $F_{s,t}$ are determined by the details of the inverse
propagator. For the Wilson action $F_s(m) = 4 \sin^2 (m/2)$, and a
more complicated function with the same continuum limit for the
Symanzik improved case. For the actions we consider below, the need to
avoid extra ghost poles in the gluonic propagator (not to be confused
with ghost fields) means that the temporal function is always unimproved:
$F_t(E) = 4 \sinh^2 (E/2)$.

Adding one loop corrections, the on-shell condition becomes
\ben
\GaR^{(\a=1)}_{\mu\nu}(\bp) = \Ga^{(\a=1)}_{\mu\nu}(iE_R,m_0,0,p_3)
- g^2 \S_{\mu\nu}(iE_0,m_0,0,p_3) = 0\;.
\label{GaR}
\een
Since we are working to $O(g^2)$, it is sufficient to evaluate
$\S_{\mu\nu}(\bp)$ at the tree level on-shell energy, $E_0$, as per
Eqn.~(\ref{eqn_tree_zero}). In general this requires taking into
account the full matrix structure of $\Ga^{(\a)}_{\mu\nu}$ but for the
form of the momentum chosen, Eqn.~(\ref{mom}), it can be shown that of
the elements $\GaR^{(\a)}_{2\,\nu}$ and $\GaR^{(\a)}_{\mu\,2}\;$, only
$\GaR^{(\a)}_{2\,2}$ is non-zero and thus that $\bp$ is a zero of this
on-diagonal element and no diagonalization of $\Gamma^{(\a)}_{\mu\nu}$
is required. For given $L, p_3$, we determine the bare pole value
$E_0$ by numerical solution.

As described in
\cite{luwe} 
the field theory for the $A$-meson is a 2D theory, and we can write by
analogy with Eqn.~(\ref{eqn_tree_zero}) an effective dispersion
relation for the infrared modes (i.e. small $p_3$) in terms of a
renormalized mass, $m_R$, and anisotropy, $\chi_R$:
\ben
{\chi_R}^2 F_t(E_R) = {p_3}^2 + F_s(m_R).
\label{eqn_ren_disp}
\een
Using Eqn.~(\ref{GaR}), we have
\ben
\frac{1}{\chi} \left( -\chi^2 F_t(E_R) + {p_3}^2 + F_s(m_0) \right)
- g^2 \Sigma(iE_0,m_0,0,p_3) = 0.
\een
Substituting in Eqn.~(\ref{eqn_ren_disp}) gives
\ben
-\frac{\chi^2}{{\chi_R}^2} \left( {p_3}^2 + F_s(m_R) \right) + 
{p_3}^2 + F_s(m_0) = g^2 \chi \Sigma(iE_0,m_0,0,p_3).
\een
We define 
\ben
Z(g^2,\chi) \equiv \chi_R/\chi = 1 + \eta(\chi) g^2 + O(g^4) \; , 
\label{eqn_z}
\een
and $m_R = m_0 + g^2 m_1 + O(g^4)$, and at one loop obtain the relation
\ben
2 \eta \left( {p_3}^2 + F_s(m_0) \right) - m_1 F_s^\prime(m_0) = 
\chi \Sigma(iE_0,m_0,0,p_3).
\label{eta_line}
\een
For $L \rightarrow\infty$ we have
\ben
\eta \bb {p_3^2 + m_0^2 \over \chi^2} \eb - {m_1 m_0 \over \chi^2} = 
{\S \over 2\chi}\;.
\een
The values of $\eta(L)$ and $m_1(L)$ are determined by a (very good)
straight line fit to Eqn.~(\ref{eta_line}).

\subsection{\label{sec_calc} The calculation}

We work with the $D=3+1$ SU(3) gauge theory and consider the Wilson
action (W) and the Symanzik improved action (SI) defined in
\cite{alea}.

\subsubsection{the Wilson action}

The Wilson action has a two dimensional coupling space, and is
\ben
S_W(\beta,\chi) = -\beta \left(
\chi \sum_{x,s} P_{s,t} + \frac{1}{\chi} \sum_{x,s>s'} P_{s,s'}
\right)
\een
where $s,s'$ run over spatial links in different, positive directions,
$P_{s,s'}$ and $P_{s,t}$ are plaquettes and $\chi$ is the
(unrenormalized) anisotropy as {\it per}
Eqn.~(\ref{eqn_anis_ac}). Spatial and temporal tadpole improvement
factors, $u_{s,t}$, arising from favourite self-consistency conditions
may be written in, but this amounts merely to a reparametrization of
the same action
\ben
\left( \beta \equiv \frac{2N}{g^2} \right) = 
\left( \frac{\beta_0}{{u_s}^3 u_t} \equiv \frac{2N}{{g_0}^2} \right) \; ,
~~~~
\chi = \frac{\chi_0 u_s}{u_t} \; ,
\label{redefs}
\een
and the measured anisotropy, $\chi_R$, is invariant. We shall evaluate
Eqn.~(\ref{eqn_z}) perturbatively.  In doing so, we span the full
space of couplings and our calculation of $Z(g^2,\chi)$ applies
equally well to any form of the Wilson action.

To one loop the diagrams for self energy function for the gluon are
shown in Fig.~\ref{prop_graphs}(a-e).  The calculations were done on
lattices with $8 L^2 I^2$ twisted momentum grid points. The loop
momentum $\bk$ to be summed over is
\bea
(k_0,k_1,k_2,k_3) & = &
({2\pi\over I}n_0,{2\pi\over 3L}n_1,{2\pi\over 3L}n_2,{2\pi\over I}n_3) \nn
\\
&& 0 \le n_0,n_3 < I ,~~0 \le n_1,n_2 < 3L\;.
\eea
There is a pole in the integrands in $\bk$ of the graphs in
Fig.~\ref{prop_graphs}(a,c) for on-shell external momentum so the
$k_0$ integration contour is shifted by $k_0 \rightarrow k_0 -
iE_0/2$. In addition, we use the change of variables $\bk \rightarrow
\bk^\prime$ suggested by L\"{u}scher and Weisz
\cite{luwe}
\ben
k^\prime_\mu = k_\mu - \a_\mu \; \sin(k_\mu)\;,
\een
which gives an integrand with much broader peaks, which is easier to
evaluate numerically. It is easy to see that a reasonable choice of
parameter is $\a_\mu \sim 1-(\chi L_\mu)^{-1}$ and this was found to work
well, significantly reducing the dependence on $L$.

The integrals were done as direct summations over the discrete
momentum modes but in principle they can be calculated using an
adaptive Monte-Carlo integrator even for finite $L,I$ . A common
example of such an importance sampling integration package is {\sc
Vegas}
\cite{lepage78}
(and described in 
\cite{numrec}), 
but this expects the integrand to be a continuous function of its
arguments. This can be achieved by a change of variable using a
stepped function $Z(x,L)$ defined by
\ben
Z(x,L) = int(xL)\;,~~~0 \le x < 1\;,
\een
so that $Z(x,L)$ takes integer values in the range $0,\ldots,L-1$ and
a given discrete momentum component is $k(x) = 2\pi Z(x,L)/L$. It
turns out that for one loop integrations the summation over modes is
much more efficient. For two loops, however, it will be necessary to
use {\sc Vegas} to get a result of acceptable accuracy.

We considered lattice sizes in the range $4 \le L \le 32$ and $I$
large enough for the error due to finite $I$ to be essentially
undetectable. In practice, $I > 50$ sufficed. We calculated the
self energy at each $L$ for a number of very small values of $p_3$ and
using Eqn.~(\ref{eta_line}) calculated the parameters $\eta(L)$ and
$m_1(L)$. All computations were done on a single processor PC and took
between 2 and 16 hours per integral, depending on $L$.

A fit as a function of $L$ allows the $L \rightarrow \infty$ values to
be deduced. In Fig.~\ref{wil_eta_L} we show the $L$-dependence of
$\eta(L)$ for $\chi=2$, and the (excellent) fit
shown is given by
\ben
\eta(L) = 0.085303 - 0.0188471 {1 \over L^2} - 0.0073487 
{\log(L^2) \over L^2}\;.
\een
For $\chi = 1$ we found that, as expected, $\eta(L) = 0$ for all $L$
and that the mass renormalisation parameter $m_1$ was
\ben
{m_1 \over m_0} = -0.037923(9)\;,
\een
a value which agrees with that given by Snippe using the background
field method
\cite{snip}. 
Extrapolated values of $\eta$ shown in Table~\ref{si_eta_bchi_vals}
agree closely with results obtained by P\'{e}rez and van Baal
\cite{peba}, 
also using background field gauge. For instance, we find 
\ben
\eta(\chi=2) = 0.085303 \; (1)\;,~~~~\eta(\chi=4) = 0.127901 \; (2)\;,
\een
compared with P\'{e}rez's and van Baal's values of 0.0853037 and
0.1278990 respectively. Combining our data with that in 
\cite{peba}, 
we find a close phenomenological fit is
\ben
\eta(\chi) = 0.1687 \; (2) - 0.16397 \; (4) \frac{1}{\chi} 
- 0.005245 \; (2) \frac{1}{\chi^2}.
\label{eqn_chi_wil}
\een

\subsubsection{the Symanzik improved action}

The general form of the action has, in addition to the couplings
$\beta$ and $\chi$, an additional parameter, $v$:
\bea
S_{SI}(\beta,\chi,v) & = &
-\beta \chi \sum_{x,s} \left\{
\frac{4}{3} P_{s,t} - \frac{1}{12}\frac{R_{ss,t}}{v^2} \right\}
\nn \\
&& -\beta \frac{1}{\chi} \sum_{x,s>s'} \left\{
\frac{5}{3} P_{s,s'} - \frac{1}{12}\frac{R_{ss,s'}}{v^2}
-\frac{1}{12}\frac{R_{s's',s}}{v^2} \right\}
\label{eqn_si_act}
\eea
where $R_{ss,s'}$ and $R_{ss,t}$ are $2\times 1$ loops. The anisotropy
renormalisation factor, $Z(\chi,\beta,v)$, is a function of
all three parameters.

It is usual to restrict the simulated parameters to a two dimensional
surface in the coupling space. The particular surface is determined by
the chosen method of tadpole improvement, demanding $v =
u_s(g^2,\chi)$ for the reparametrization in Eqn.~(\ref{redefs}). If,
for given anisotropy, $u_s$ has a perturbative expansion $u_s(g^2) = 1
+ d_s g^2 + O(g^4)$, then we can rewrite the action as
\bea
S_{SI}(\beta,\chi,v) & = & S_{SI}(\beta,\chi,v=1)
+ g^2 \D S_{SI} + O(g^4) \; ,
\nn \\
\D S_{SI} & = & - \beta d_s \sum_{x,s>s'} - \frac{1}{6} \left\{
\chi R_{ss,t} + \frac{1}{\chi} \left( R_{ss,s'}+R_{s's',s} \right)
\right \}
\; .
\eea
This form is numerically more convenient than the reparametrized form,
in that the dependence of the action on the gauge coupling is lessened.

In this paper we calculate $Z(g^2,\chi)$ over two such surfaces:
$v=1$, corresponding to an action without tadpole improvement (in
which case $\Delta S_{SI}$ makes no contribution), and the surface
$v=u_s$ corresponding to Landau mean link improvement where $u_{s,t}$
are defined as the expectation values of the traced link matrices in
the Landau gauge
\cite{lepa0} 
\ben
u_s = \la U_s\ra_L\;,~~~~u_t = \la U_t\ra_L\;.
\een
$U_{s,t}$ are gauge fields in the spatial and temporal directions. The
actual values of $u_{s,t}$ are established self-consistently by an
appropriate iteration procedure, usually for fixed~$\beta_0$
and~$\chi_0$ (which we comment upon in Section~\ref{sec_summary}). As
$\Delta S_{SI}$ is $O(g^2)$ relative to other terms in the actions, we
treat it as a counterterm, giving rise to a gluon propagator insertion
shown in Fig.~\ref{prop_graphs}(f). The coupling, $d_s$, is determined
numerically in each case for an identical lattice
\cite{lat02,meanlink02}.

In both cases, the actual calculation proceeds as for the Wilson
action (the only changes in the loop integration code being the {\sc
Python}--generated input files). In Fig.~\ref{si_eta_bchi} we plot
$\eta(\chi)$ as a function of $\chi$. The results are also presented
in Table~\ref{si_eta_bchi_vals}. The slightly larger errors in the
full calculation reflects a greater uncertainty in the extrapolation
in~$L$ due to reduced data.

These values can be fitted extremely well, as is demonstrated in
Fig.~\ref{si_eta_bchi}, by
\ben
\eta(\chi) = 0.0602 \; (1) - 0.0656 \; (2) \frac{1}{\chi} 
- 0.0237 \; (1) \frac{1}{\chi^2}
\label{eqn_chi_noct}
\een
without the mean link counterterm contribution, and 
\ben
\eta(\chi) = 0.0955 \; (4) - 0.0702 \; (16) \frac{1}{\chi} 
- 0.0399 \; (14) \frac{1}{\chi^2}
\label{eqn_chi_ct}
\een
when this is included.

\subsection{\label{sec_comparison}comparison with measured data}

The goal of this paper is to provide determinations of the
renormalied anisotropy using perturbation theory. It is clear that
for sufficiently small couplings the one loop result will be
sufficient, but at couplings more typical of current lattice
simulations we must check what systematic truncation errors are
introduced by neglecting contributions of higher orders. We can do this
by comparing the perturbative predictions with what measurements have
been made in Monte Carlo simulations.

We start with the Symanzik improved action, for which measurements of
$\chi_R$ have been published in
\cite{alea,shigemitsu02},
derived from torelon dispersion relations and the so-called sideways
potential. 

We have obtained as a perturbative series, $Z(g^2,\chi,v) = 1 +
\eta(\chi,v) g^2 + O(g^4)$. Such a series, however, shows poor
convergence and thus large truncation errors. We can re-express it as
a power series in the `self-consistent', or `boosted' coupling,
${g_0}^2$:
\ben
Z({g_0}^2,\chi) = 1 + \eta(\chi,v=u_s) {g_0}^2 + O({g_0}^4) \; ,
\een
using the definitions in Eqn.~(\ref{redefs}).  For appropriate choices
of tadpole improvement schemes, the convergence is much improved. In
resumming the series, we formally require only that $u_{s,t}$ are
self-consistent for fixed $\beta_0,\chi_0$ up to a suitable order of
perturbation theory, and that $v=u_s$ to the same order. There are
arguments, however, for satisfying these conditions at a numerical
level for greater convergence. The one loop result is, of course,
unchanged under resummation.

The action in
\cite{alea,shigemitsu02}
contains explicit factors of $u_{s,t}$ and before applying $Z$ it is
necessary to rescale $\chi_0$ to $\chi$ using the simulated values of
the mean link factors. In Table~\ref{tab_meas_chi} we compare the
one loop perturbative determination with the published
measurements. The couplings are larger than those used in many
simulations (giving a relatively coarse spatial lattice), and the
anisotropy factors are at the higher end of those usually employed,
making this a rather stringent test of the perturbative
series. Despite this, we find this error to be uniformly small. For
$\chi_0 \le 4$ the error is 2\%, and for $\chi_0=6$ it is only 3\%. If
we attribute the difference between the measured and one loop estimate
to be that due to truncation of the perturbative series, it is clear
that the two loop contribution to $Z$ is very small. The range of
couplings, however, is insufficient to allow us to reliably estimate
the two loop coefficient from the data.

A comparison of the one loop perturbative and measured anisotropy has
been carried out for the Wilson action by Klassen
\cite{klassen98}.
Corroboration of this analysis is hampered by a lack of published
tadpole improvement factors. Whilst $\chi$ is quoted directly,
$u_{s,t}$ are required to obtain the self-consistent coupling,
${g_0}^2$, appropriate to simulated values of $\beta$.  We may
estimate the tadpole factors, but in doing so we introduce a second
systematic error into the estimate of $\chi_R$. This arises from the
difference between our estimate and the numerically self-consistent
tadpole values. It is in addition to the systematic error arising from
the one loop truncation of the perturbative expansion of $Z$. We
anticipate the latter being of order 3\%, and we thus require values
of $u_{s,t}$ at least this accurate. Such an estimate can come from a
two loop perturbative estimate
\cite{lat02,meanlink02}.
Using this we have confirmed using Landau mean link improvement
Klassen's analysis (which used plaquette tadpole factors). We find the
combined two loop tadpole, one loop anisotropy prediction for $\chi_R$
to be correct to within 5\% for $\beta \ge 5.6$, and to 2\% for $\beta
\ge 6.3$ for $1.5 \le \chi \le 6$. With the correct $u_{s,t}$ values,
a more precise estimate of the truncation errors in $Z(\chi,g^2)$
would be possible, and we expect them to show that the one loop
anisotropy renormalisation calculation is as accurate for the Wilson
action as we have demonstrated it to be for the Symanzik improved
case.

\section{\label{sec_summary}Summary}

We have presented perturbative calculations of the renormalisation of
the anisotropy in simple gluonic lattice actions for SU(3) in
$D=3+1$. These have been carried out by studying the infrared
dispersion relation of the on-shell gluonic propagator, regulated
through the use of twisted boundary conditions as in
\cite{luwe}.
We have reviewed the derivation of the Feynman rules for general
actions and anisotropy structure, using both twisted and untwisted
boundary conditions. We have (briefly) outlined our method for
generating the gluon $r$-point vertex functions using a {\sc Python}
code and evaluating the loop integrals using a compiled programming
language. The strength of this dual--pronged approach is the
generality; at the price of not carrying out optimisation specific to
a particular action, a flexibility to rapidly change actions is
acquired. Further details of the implementation will be given in
\cite{meanlink02}.

We have focussed our attention on two actions, the Wilson plaquette
action and an action (Symanzik) improved at tree level through the
addition of $1 \times 2$ loops extending two lattice spacings in
spatial directions only. This lack of symmetry is necessary to avoid
spurious ghost poles in the gluonic propagator, and in consequence
even an isotropic lattice will not remain so under quantum
corrections. Temporal improvement is not crucial, of course, when
$a_t$ is small.

Both actions are considered with and without tadpole improvement of
the links. In the case of the Wilson action no explicit reference to
$u_{s,t}$ need be made in the perturbation theory. Specific tadpole
improvement schemes manifest only in rescaling the gauge coupling.

For the Symanzik improved action, the lack of temporal improvement
means all reference to $u_t$ can be reparametrized away, but
dependence on $u_s$ remains in the form of a counterterm to the
(tadpole) unimproved action. The presence of the counterterm
necessarily specialized the perturbative calculation to one form of
tadpole improvement, and we selected self-consistent Landau gauge mean
links for this work.

There are also advantages in using this reparametrization when tuning
actions such as $S_{SI}$ for Monte Carlo simulation, when explicit
tadpole factors $u_{s,t}$ are often present. reparametrizing to remove
dependence on $u_t$ makes the finding of a self-consistent $u_s$ for
fixed $(\beta,\chi)$ a one parameter problem. Similarly tuning $u_t$
is done by varying $(\beta_0,\chi_0,u_t)$ together along the single
parameter curve of constant $(\beta,\chi,u_s)$. This two stage
approach is much faster than a simultaneous exploration of the two
parameter space.

For the Symanzik improved action, we presented results over a spread
of bare anisotropies $1 \le \chi_0 \le 8$, and give interpolating
fits, Eqns.~(\ref{eqn_chi_noct},~\ref{eqn_chi_ct}), respectively with
and without (Landau) mean link tadpole improvement. Comparing these
numbers with measurements in simulations shows that the one loop
perturbative determinations of the renormalized anisotropy are
accurate to within 2\% for $\chi \le 4$ across a range in couplings
and lattice spacings much larger than would typically be used in
lattice simulations. For $\chi=6$, which is larger than that currently
employed in large scale simulations, the deviation is 3\%.

In the case of the Wilson action we have verified existing results
(calculated using the background field method) by Snippe
\cite{snip}
and P\'erez and van Baal
\cite{peba}.
Using an interpolating fit, Eqn.~(\ref{eqn_chi_wil}), we have carried
out a comparison with lattice simulation results by Klassen
\cite{klassen98}.
We find the agreement to be approximately 3--4\% for gauge couplings
in the typically simulated range and $\chi \le 6$, but this
uncertainty includes a systematic error from having to estimate
tadpole improvement factors, and the error in the anisotropy alone is
almost certainly less.

We conclude that the renormalisation of the anisotropy, which is
difficult to calculate accurately in simulations, is well described by
lattice perturbation theory. If the anisotropy is to be used, for
instance, to set the scale in lattice calculations as per
Eqn.~(\ref{eqn_setscale}), then a 3\% systematic error in $\chi_R$
is sufficiently small that is unlikely to represent the dominant
uncertainty in the final estimate, and the one loop determinations
presented in this paper are all that is required.

\begin{acknowledgments}

We are pleased to acknowledge the use of the Hitachi SR2201 at the
University of Tokyo Computing Centre and the Cambridge--Cranfield High
Performance Computing Facility for this work.

\end{acknowledgments}

\newpage

%
%
%
%
%

\btab[p]
\bec
\btabu{|l|c|l|l|}\hline
& Wilson & \multicolumn{2}{c|}{Spatially Improved} \\ 
\cline{3-4}
$\chi$ & & \multicolumn{1}{c|}{(a)} & \multicolumn{1}{c|}{(b)} \\ 
\hline
1   & 0           & $-0$.02907 (1) & $-0$.01461 (30) \\
1.5 & -           &  0.005654 (5)  & 0.03163 (30) \\
2   & 0.085303 (1) &  0.021446 (4) & 0.05037 (30) \\
2.5 & -           &  0.03027 (1)   & 0.06097 (30) \\
3   & -           &  0.035692 (8)  & 0.06753 (30) \\
3.5 & -           &  0.03955 (1)   & 0.07218 (30) \\
4   & 0.127901 (2) &  0.04251 (1)  & 0.07567 (30) \\
5   & -           &  0.04851 (2)   & 0.08000 (30) \\
6   & -           &  0.04851 (2)   & 0.08279 (30) \\
8   & -           &  0.05172 (5)   & 0.08637 (30) \\
\hline
\etabu
\enc
\caption{\label{si_eta_bchi_vals}\small Anisotropy renormalisation,
$\eta$, for various $\chi$ for Wilson and Symanzik improved
actions. The latter is shown (a) without and (b) with the spatial mean
link counterterm contribution of Fig.~\ref{prop_graphs}(f).}  
\etab

\btab[p]
\bec
\btabu{|l|l|l|l|l|l||l|l|l|}
\hline
$\beta_0$ & $\chi_0$ & 
\multicolumn{1}{c|}{$u_s$} & 
\multicolumn{1}{c|}{$u_t$} & 
\multicolumn{1}{c|}{$a_s^{-1}$} & 
\multicolumn{1}{c||}{$\chi_R^{\rm (meas.)}$} &
\multicolumn{1}{c|}{$\chi$} &
\multicolumn{1}{c|}{$\eta(\chi)$} &
\multicolumn{1}{c|}{$\chi_R^{\rm (pert.)}$} \\
&&&& (MeV) &&&& \\
\hline
1.7 &  4 & 0.7164 & 0.98295 & 661 (11) & 
3.56 (2) & 2.915 &  0.0667 &   3.601 \\
    &  6 & 0.7100 & 0.99158 & 779 (28) &
5.28 (2) & 4.296 &  0.0770 &   5.463 \\
\hline
1.8 &  4 & 0.7279 & 0.98243 & 797 (21) & 
3.61 (2) & 2.964 &  0.0673 &   3.629 \\ 
    &  6 & 0.7216 & 0.99208 & 839 (9) & 
5.31 (2) & 4.364 &  0.0773 &   5.489 \\
\hline
2.4 &  3 & 0.7868 & 0.9771 & 1200 (50) & 
2.71 (3) & 2.416 &  0.0596 &   2.776 \\
\hline
\etabu
\enc
\caption{\label{tab_meas_chi}\small Renormalized anisotropy as
determined from torelon dispersion relations and the sideways
potential
\cite{shigemitsu02,alea},
and from one loop perturbation theory.}
\etab

%
%
%
%
%

\bef
\bec
\epsfig{file=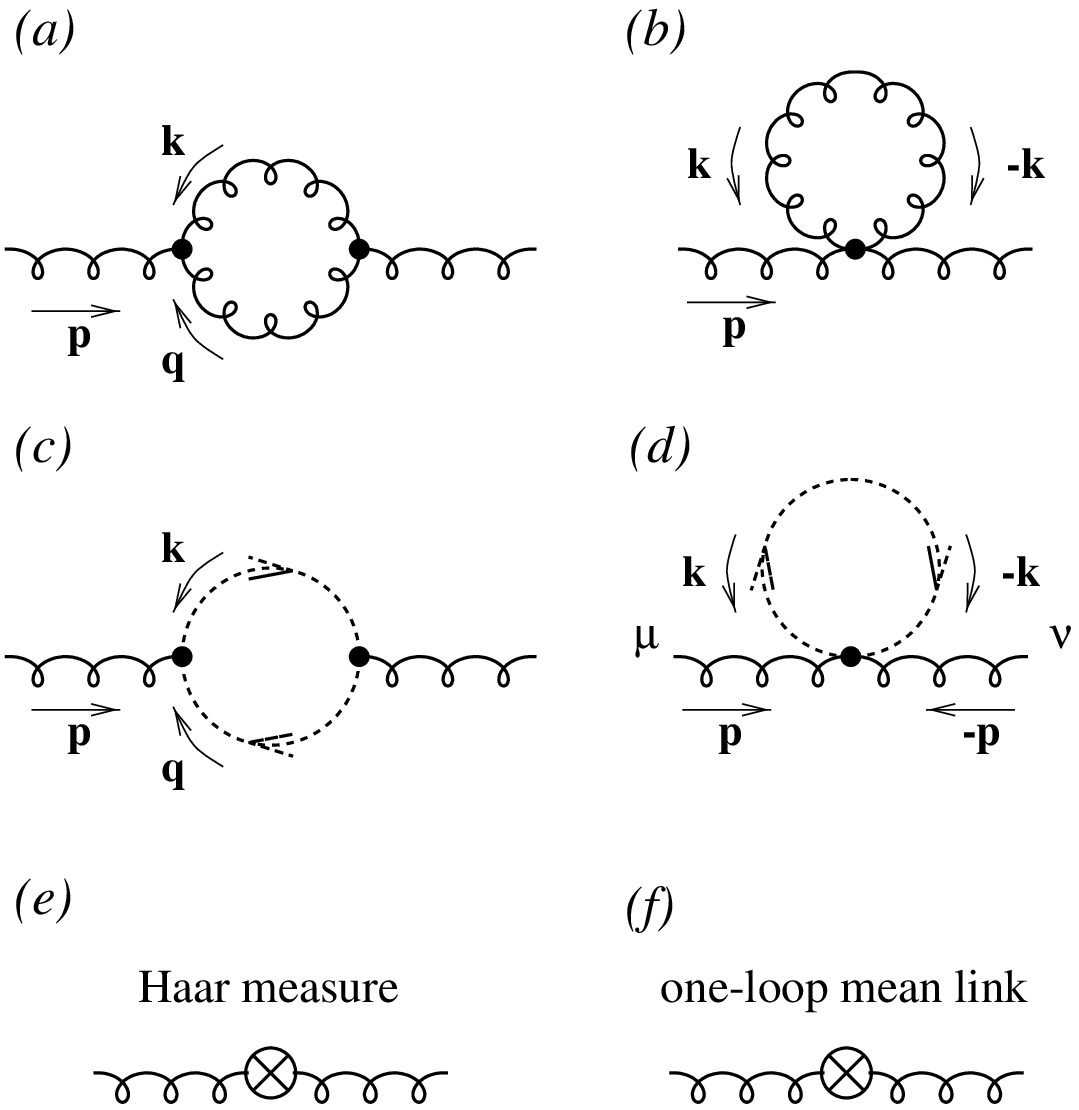,height=110truemm}
\enc
\caption{\label{prop_graphs}\small Self energy graphs at $O(g^2)$ for
the gluon propagator. Graph (f) arises from treating the spatial mean
link factor as a counterterm, and has a coupling proportional to the
one loop expression for this.}  
\enf

\bef
\bec
\epsfig{file=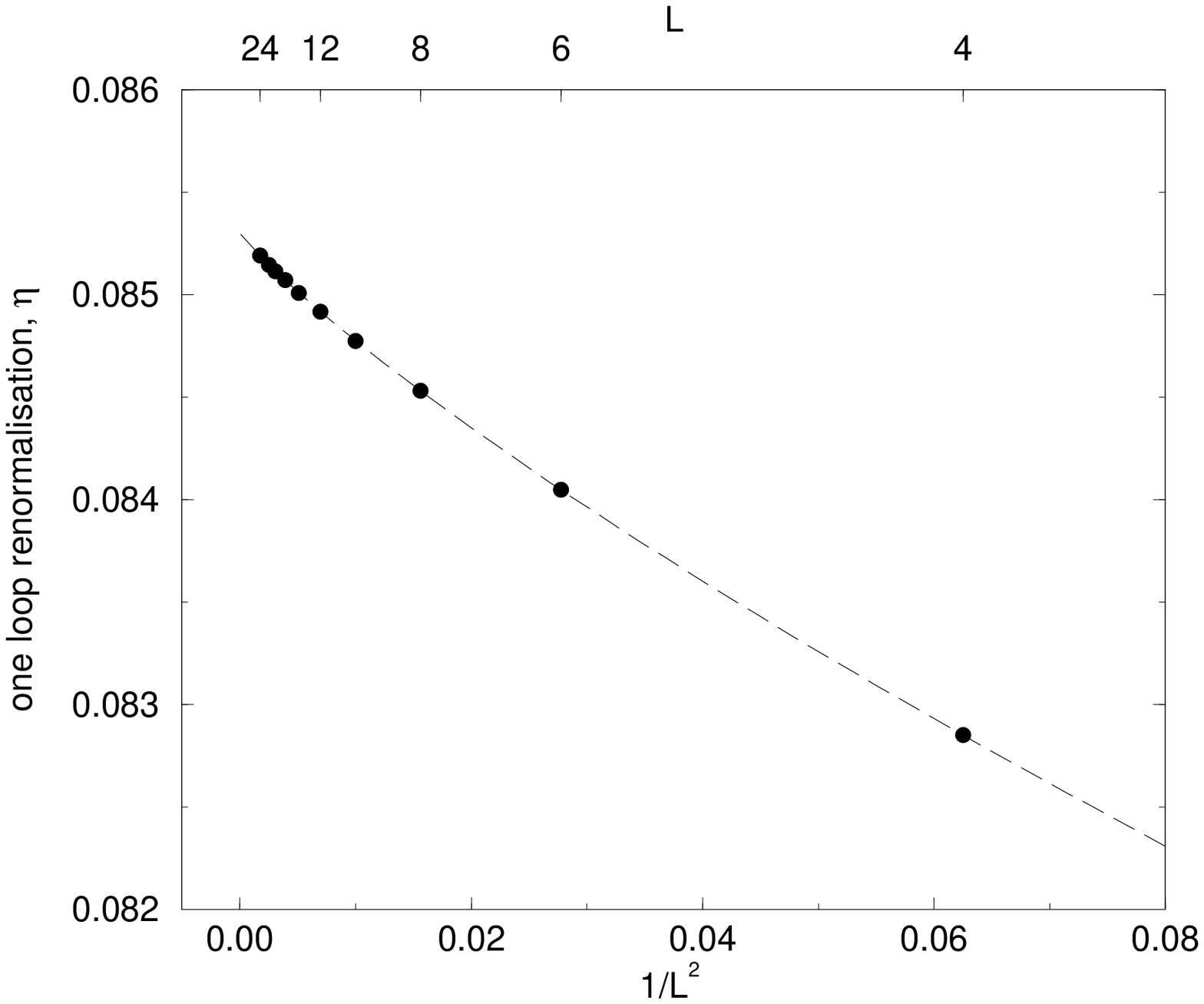,height=110truemm}
\enc
\caption{\label{wil_eta_L}\small
Anisotropy renormalisation for the Wilson action as a function
of $L$ for bare anisotropy $\chi=2$.}
\enf

\bef
\bec
\epsfig{file=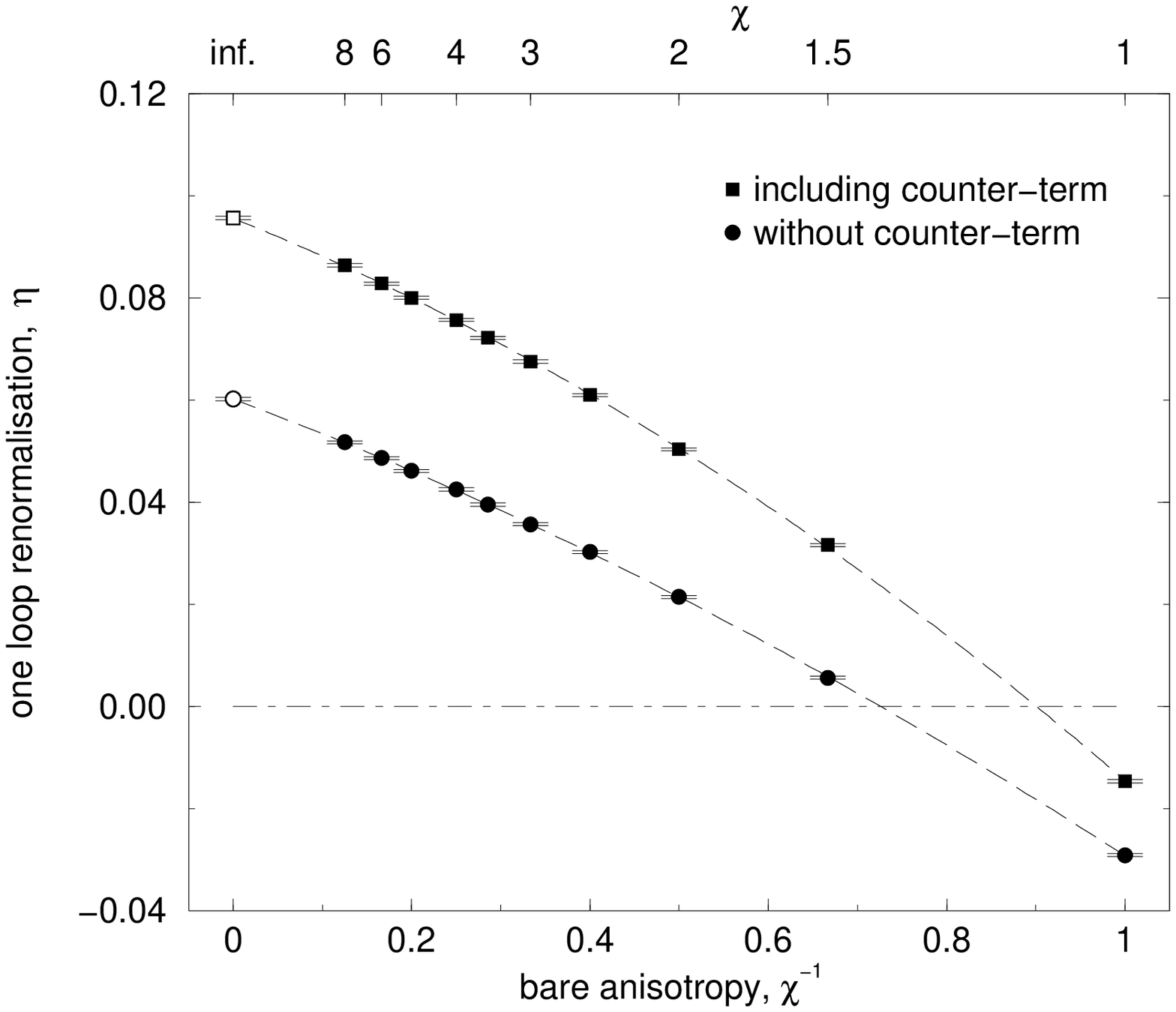,height=110truemm}
\enc
\caption{\label{si_eta_bchi}\small
Anisotropy renormalisation for the spatially improved action as a function
of the bare anisotropy $\chi$, showing fits quadratic in $\chi^{-1}$.}
\enf

\end{document}